\documentclass[final]{aipproc}

\layoutstyle{6x9}

\input psfig.sty

\def\ltsima{$\; \buildrel < \over \sim \;$}
\def\lsim{\lower.5ex\hbox{\ltsima}}
\def\gtsima{$\; \buildrel > \over \sim \;$}
\def\gsim{\lower.5ex\hbox{\gtsima}}

\begin{document}

\title{Observational Gamma-ray Cosmology}

\author{Joel R. Primack}{
  address={Physics Department, University of California, Santa Cruz,
CA 95064 USA}}

\author{James S. Bullock}{
  address={Department of Physics and Astronomy, University of
California, Irvine, CA 92697 USA}}

\author{Rachel S. Somerville}{
  address={Space Telescope Science Institute, 3700 San Martin Drive
Baltimore, MD 21218 USA}}

\begin{abstract}
We discuss how measurements of the absorption of
gamma-rays from GeV to TeV energies via pair production on the
extragalactic background light (EBL) can probe important issues in
galaxy formation.  Semi-analytic models (SAMs) of galaxy formation,
based on the flat LCDM hierarchical structure formation scenario, are
used to make predictions of the EBL from 0.1 to 1000 microns.  SAMs
incorporate simplified physical treatments of the key processes of
galaxy formation -- including gravitational collapse and merging of
dark matter halos, gas cooling and dissipation, star formation,
supernova feedback and metal production.  We will summarize SAM
successes and failures in accounting for observations at low and high
redshift.  New ground- and space-based gamma ray telescopes will help
to determine the EBL, and also help to explain its origin by
constraining some of the most uncertain features of galaxy formation
theory, including the stellar initial mass function, the history of
star formation, and the reprocessing of light by dust.  On a separate
topic concerning gamma ray cosmology, we discuss a new theoretical 
insight into the distribution of dark matter at the center of the
Milky Way, and its implications concerning the high energy gamma rays
observed from the Galactic center.
\end{abstract}

\maketitle


\section{Introduction}

The main process that removes high energy gamma rays enroute from
remote active galactic nuclei (AGN) to our detectors is absorption via
$\gamma \gamma {\rightarrow} e^+ e^-$ as the gamma rays move through
the evolving extragalactic background light (EBL).

\begin{figure}[h]
  \psfig{file=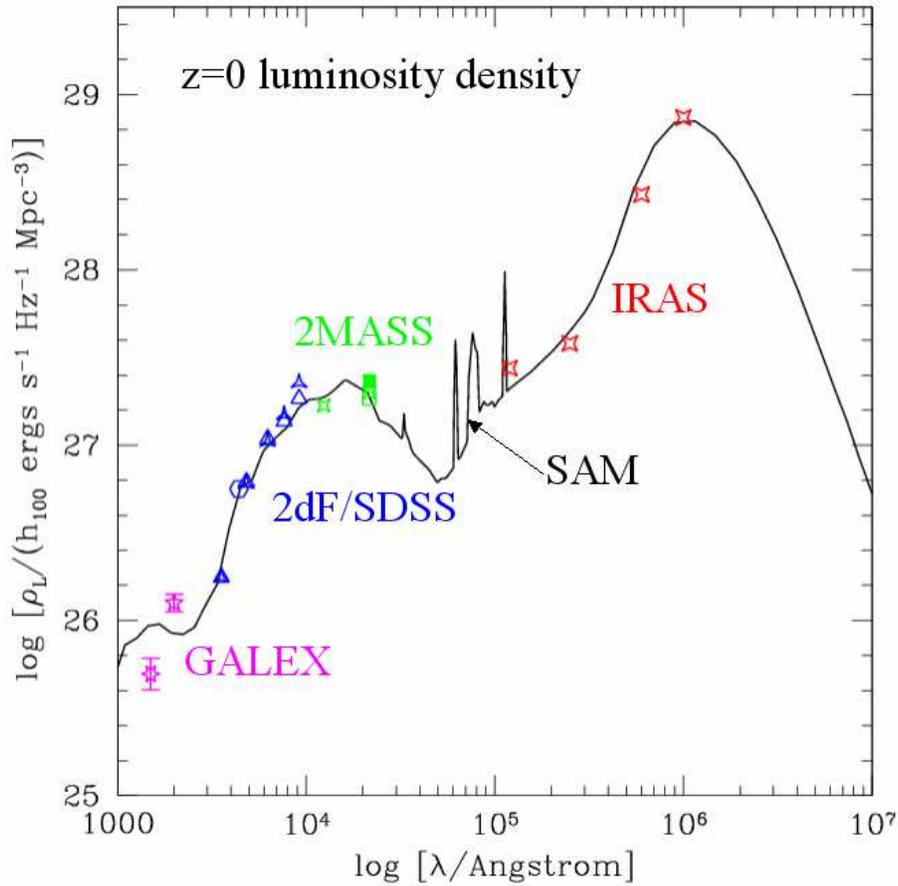,height=12cm}
  \caption{Luminosity Density vs. Wavelength in the Nearby Universe.
  The black curve is the prediction of our current Semi-Analytic Model
  (SAM); the points are observational data.}
\end{figure}
\begin{figure}[ht]
  \includegraphics[height=10cm]{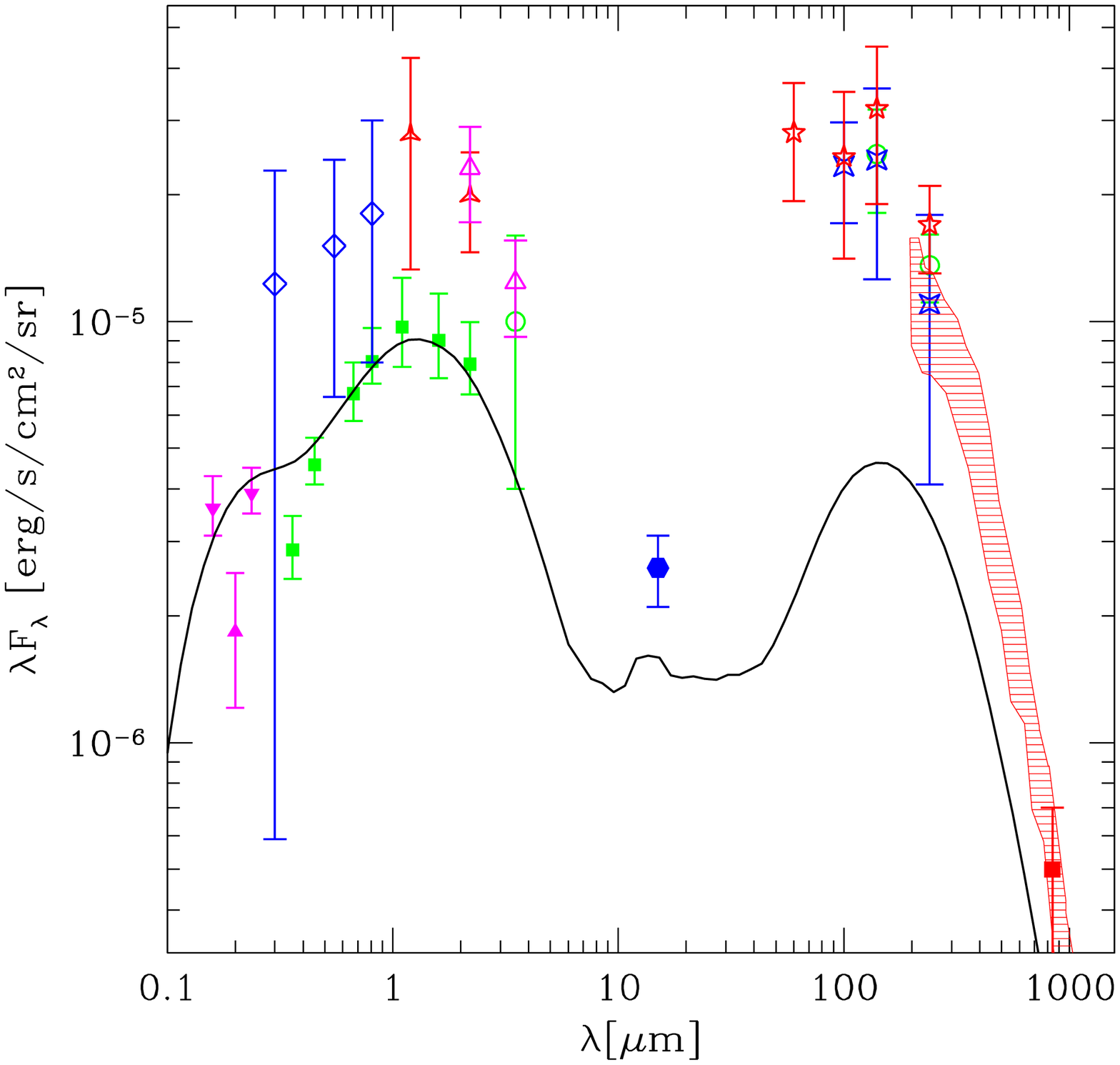} \caption{Predicted
  Extragalactic Background Light (EBL) Compared to Observations.  The
  black curve is the prediction of the SAM discussed in the present
  paper.  Also shown are a selection of relevant observations.  The
  filled squares are lower limits from the Hubble Deep Field and other
  sources; the filled hexagon is a lower limit from ISOCAM (Elbaz et
  al. 2002); the measurements at 140 and 240 $\mu$m are from the DIRBE
  instrument on the COBE satellite, and the shaded region on the right
  of the figure is from the FIRAS instrument on COBE.}
\end{figure}

From the earliest work on the cold dark matter (CDM) theory of
structure formation in the universe (Blumenthal et al. 1984),
semianalytic models (SAMs) have been used to estimate the properties
of the population of galaxies that will form in a given cosmological
model.  These models typically treat galaxy formation and evolution
based on simple approximations such as spherical collapse, but they do
treat accurately the most secure aspects of the theory, in particular
the CDM power spectrum of fluctuations.  As dark matter theory has
progressed, it became possible to base the calculation of galaxy
formation on the approximate merging history of a population of dark
matter halos calculated using Monte Carlo methods (White \& Frenk
1991, Somerville \& Kolatt 1999), and this has been the main method
used in SAMs (e.g. Kauffmann, White, \& Guiderdoni 1993; Somerville \&
Primack 1999; Cole et al. 2000).\footnote{Modern high-resolution
simulations permit the calculation of structural merger trees which
characterize the radial profiles and angular momenta of the merging
dark matter halos (e.g. Wechsler et al. 2002), and we are now doing
SAMs based on this.  This is important for predicting the dependence
of galaxy properties on environment, but the results for large galaxy
populations are similar to those of the simpler Monte Carlo
calculations used here.}  We calculated the emission of EBL by the
entire evolving galaxy population initially using even simpler
Press-Schechter (1974) models (MacMinn \& Primack 1996), and then
using state-of-the-art SAMs and taking into account the effects of
dust, which obscures light emitted in starbursts driven by galaxy
mergers, and reradiats the energy in the far infrared (Primack,
Bullock, Somerville, \& MacMinn 1999; Primack, Somerville, Bullock, \&
Devriendt 2001).  In earlier work, it was necessary to consider
several possible cosmologies, but the fundamental cosmological
parameters have now been determined with remarkable accuracy based on
the cosmic microwave background and the large scale distribution of
galaxies at low redshift and of large low-density gas clouds of gas at
redshift $z\sim3$ \footnote{These gas clouds are responsible for what
is known as the Lyman alpha forest of absorption lines in quasar
spectra.}  (e.g. Spergel et al. 2003, Tegmark et al. 2004, Seljak et
al. 2004), giving results that are compatible with all other available
cosmological data (reviewed in Primack 2004).  As a result, in the
present paper we just consider the now-standard LCDM cosmology with
$\Omega_{matter}=0.3$, $\Omega_\Lambda=0.7$, Hubble parameter $h=0.7$,
and normalization $\sigma_8=0.9$.\footnote{Of these parameters, the
only one uncertain enough to be a concern is $\sigma_8$.  We have
adopted $\sigma_8=0.9$ here because this leads to the early ionization
of the universe indicated by the WMAP large-angle polarization, which
would require exotic sources if the value of $\sigma_8$ were
significantly lower (see e.g. Somerville, Bullock, \& Livio 2003).}

\section{Implications for Gamma-ray Attenuation}

\begin{figure}[ht]
\psfig{file=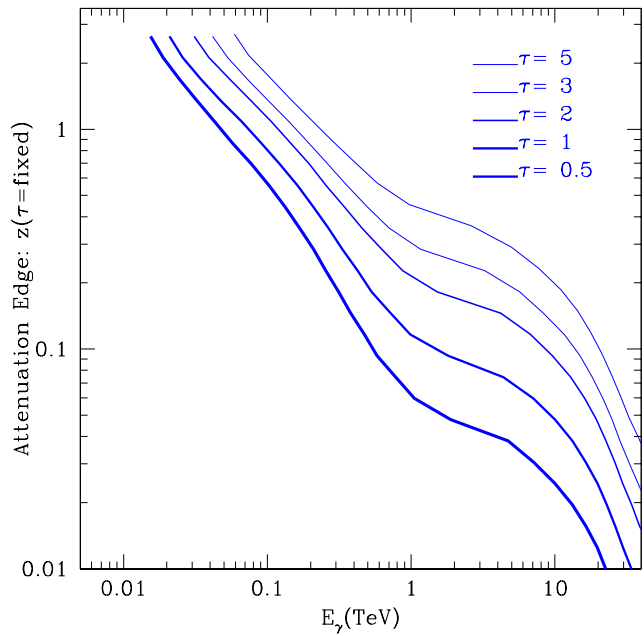,width=3.5in}
  \caption{Gamma-ray Attenuation Edge.  Each curve shows the redshift
  where the predicted attenuation as a function of gamma-ray energy is
  $e^{-1/2}$, $e^{-1}$, $e^{-2}$, $e^{-3}$, and $e^{-5}$.}
\end{figure}
\begin{figure}[h]
  \includegraphics[height=10cm]{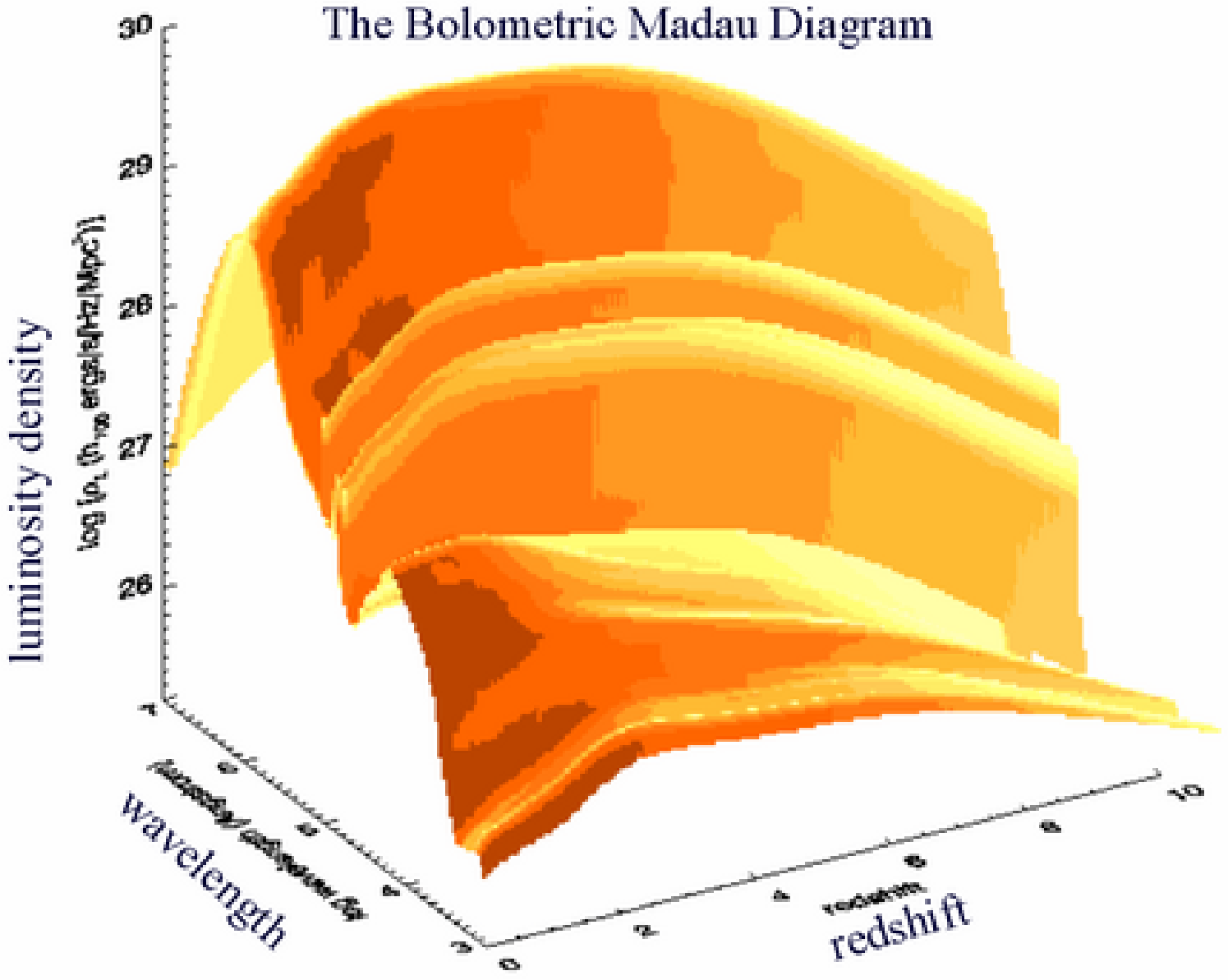}
  \caption{The Evolving Radiation Field.}
\end{figure}

In addition to the greatly increased precision and confidence in the
cosmological parameters, the main thing that has changed in
calculating the EBL from SAMs is our knowledge about the luminosity
function of galaxies, i.e. the number density of galaxies as a
function of their luminosity.  As a result of the agreement of the
three large surveys of the nearby universe, 2MASS (Cole et al. 2001),
2dF (Norberg et al. 2002), and SDSS (Blanton et al. 2003), we now know
the local luminosity density with unprecedented precision in optical
and near-infrared wavebands -- see Figure 1.  The key parameters in
SAMs, those that govern the rate of star formation and of supernova
energy feedback and metallicity yield, are adjusted to fit local
galaxy data.  The fact that this can be done well is shown by the good
agreement between the curves in Figure 1 (predictions from our current
favorite SAM) and the data from the UV to the far-IR.  However, the
optical luminosity density is lower than the best estimates of a few
years ago, which is the main reason that our EBL curves in Figure 2
have come down across the spectrum compared to our earlier estimates.

\begin{figure}[ht]
  \includegraphics[height=10cm]{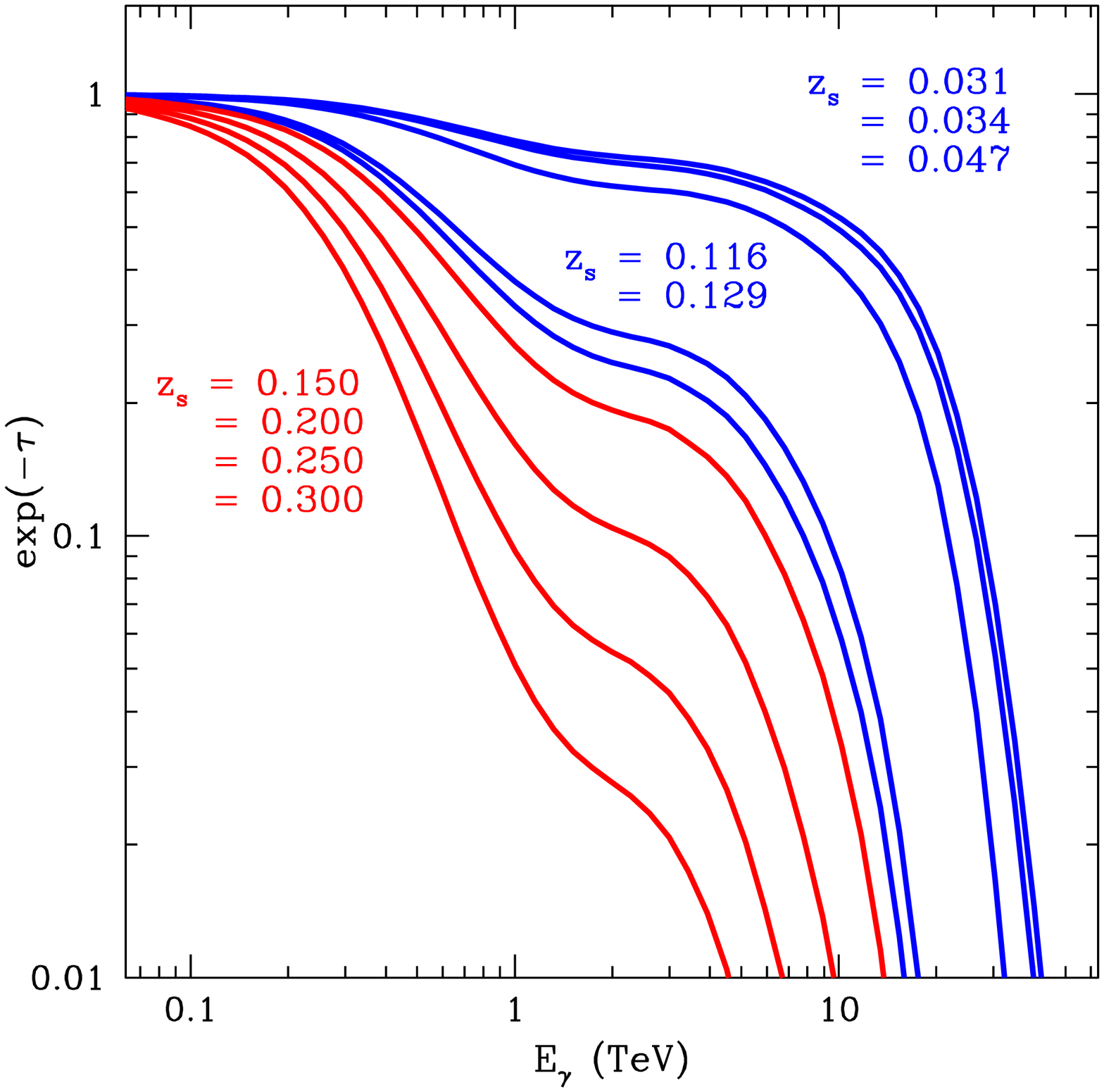}
  \caption{Gamma-ray Attenuation vs. Gamma-ray Energy.  The predicted
  attenuation is shown by the upper curves for sources at the 
  redshifts of well studied blazars, and by the lower curves for
  sources at higher redshifts.}
\end{figure}

Because the data now determine the local luminosity density in the
optical band rather precisely, and because our collisional starburst
SAM appears to do well at accounting for galaxy observations at
optical wavelengths at higher redshifts (Somerville, Primack, Faber
2001, whose predictions were shown to agree with the GOODS survey data
out to redshift $z=6$ by Giavalisco et al. 2004), we think that the
EBL in Figure 2 is probably pretty reliable in the optical and
near-IR.\footnote{The main uncertainty that we are aware of concerns
the high near-IR (1-3 $\mu$m detections shown as upward pointing
triangles in Figure 2.  We are implicitly discounting these detections
(e.g. Matsumoto et al. 2000, Wright 2004) because the systematic
errors may not be fully represented by the error bars shown, and there
is no known nearby source.  However, an interesting proposed source is
redshifted Lyman $\alpha$ radiation from early galaxies at redshifts
$z\sim 9-13$ whose shorter wavelength radiation may have been
responsible for the early ionization of the universe indicated by the
WMAP polarization data (Salvaterra \& Ferrara 2003, Magliocchetti et
al. 2003).  It may be possible to test this idea with new near-IR data
(cf. Cooray et al. 2004) and also through the predicted increased
absorption of $\sim 1$ TeV gamma rays from distant blazars such as 1ES
1426+428, which may disfavor it (cf. Costamante et al. 2004b).}  The
fact that it is somewhat lower is therefore good news for the new
generation of low energy threshold gamma ray detectors, including
HESS, MAGIC, CANGAROO III and VERITAS atmospheric Cherenkov telescopes
(ACTs), and the AGILE and GLAST gamma-ray satellites which are
scheduled to be launched in 2005 and 2007.  As gamma-ray telescopes
look at lower energy gamma rays, they can see out to higher redshifts
with less attenuation than previously predicted.  This is shown in
Figure 3, which shows curves representing varying degrees of
attenuation (in powers of $e$) predicted for gamma rays travelling
through the evolving radiation field corresponding to the EBL from our
current SAM outputs, shown in Figure 4.

Figure 5 shows the implications of our current models for attenuation
of the five well-characterized extragalactic gamma-ray souces, whose
redshifts range from $z=0.031$ to $z=0.129$.  The predicted and
observed attenuation is rather mild for the nearest sources, Mrk 421
and Mrk 501 at $z=0.031$ and 0.034.  It is only for the most distant
TeV blazar H1426+428 at $z=0.129$ that the observations (Aharonian et
al. 2003, Costamante et al. 2004a) appear to support the shape of the
predicted attenuation.  Comparing available data on four blazars with
the models, Costamante et al. (2004b) conclude that the observed
attenuation is generally consistent with our theoretical expectations.
Of course data with smaller error bars would be most welcome, and this
should be forthcoming soon with the larger and more sensitive new ACTs
now coming into operation.

\section{Theoretical Successes and Failures}
\begin{figure}[h]
  \includegraphics[height=10cm]{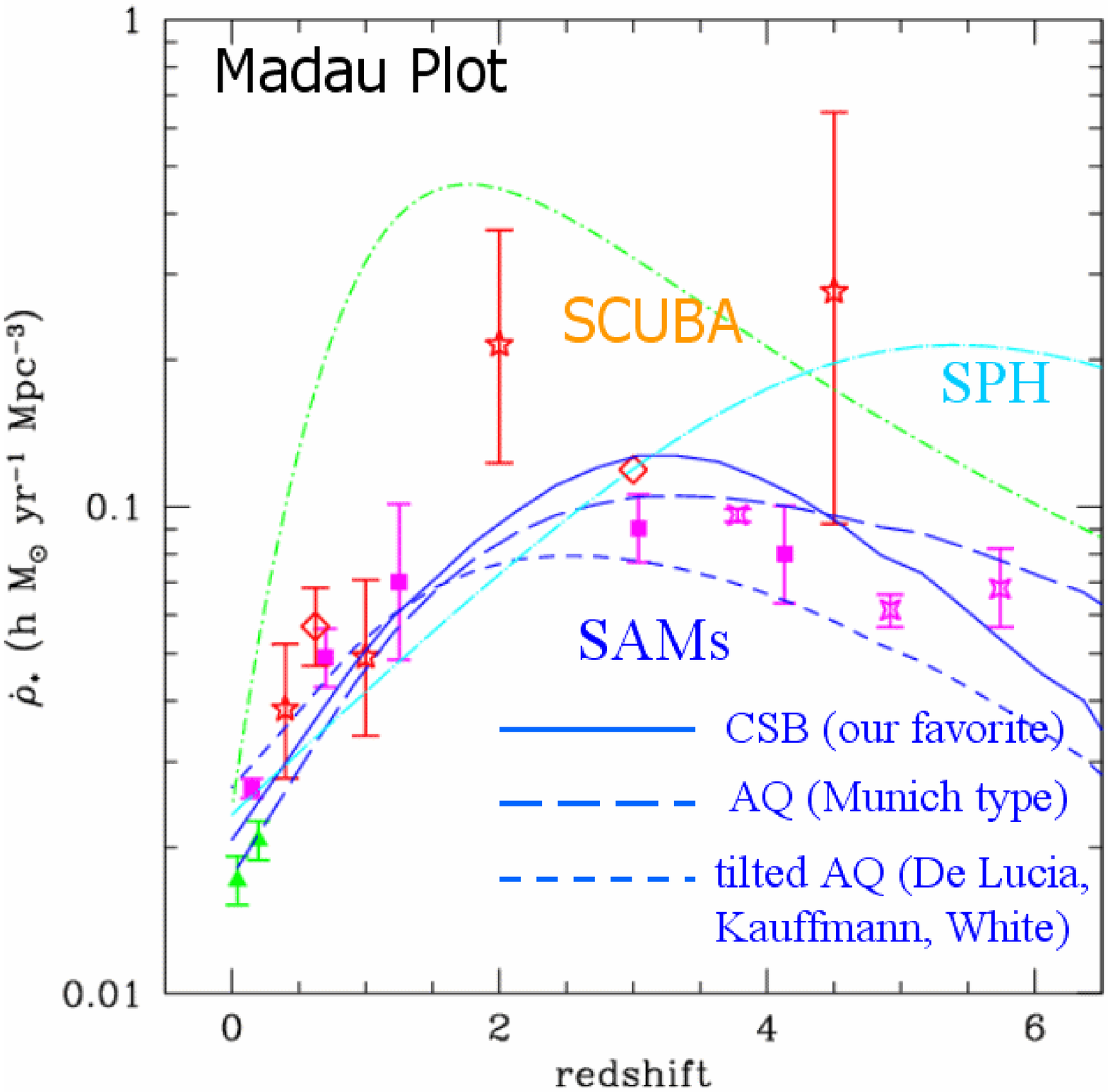} \caption{Density
  of Star Formation vs. Redshift.  The optical data points agree
  rather well with the three sorts of SAMs whose predicted Madau Plots
  are shown, collisional starburst (CSB), accelerated quiescent (AQ)
  (for both see Somerville, Primack, \& Faber 2001), and tilted AQ
  models (of the sort considered e.g. by De Lucia, Kauffmann, \& White
  2004, where the star formation rate depends on both the dynamical
  time and the circular velocity).  The curve labeled SPH shows the
  results from recent hydrodynamic simulations by  Hernquist \& Springel
  (2003).  But the two highest data points, from 850 $\mu$m
  observations by the SCUBA instrument, suggest instead a Madau plot
  like that shown by the dot-dash curve (based on a model due to
  A. Blain et al. 1999).  [This figure is based on Fig. 2 of
  Somerville 2004.]}
\end{figure}
\begin{figure}[h]
  \includegraphics[height=10cm]{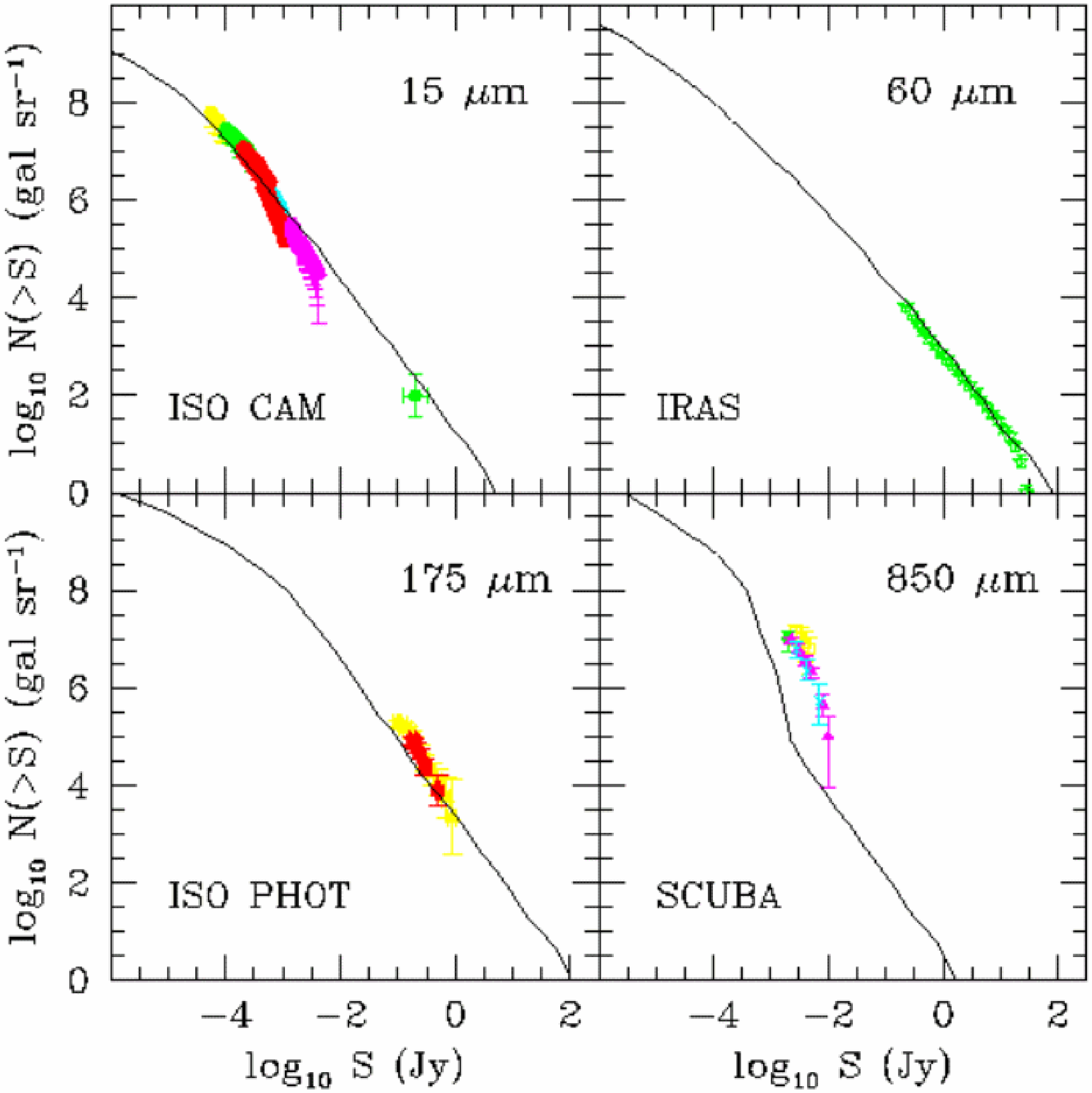}
  \caption{IR Number Counts: Predictions and Data.}
\end{figure}

There are two basic approaches to predicting the EBL, and thereby
predicting gamma-ray attenuation as a function of source redshift and
gamma-ray energy.  The old approach of ``backward models,'' still
followed by some, has been to start with the existing galaxy
population and to model the luminosity evolution of these galaxies
backward in time.  There are great difficulties in principle with this
approach.  The EBL data shown in Fig. 2 indicate that the total
energy in far-IR radiation is comparable to that in the optical and
near-IR.  However, the Milky Way, like most nearby galaxies, radiates
much more of its energy in the optical and near-IR than the far-IR.
Since most of the far-IR light must therefore have been radiated at
higher redshifts, it has been diluted by a factor of $(1+z_{\rm
emission})^{-1}$ by the expansion of the universe.  It follows that
radiation by higher-redshift galaxies must have been very different
from that of nearby galaxies, with a much larger fraction of the light
emitted in the far-IR.  Backward evolution models are further
compromised by the likelihood that such intense far-IR emission has
often been triggered by galaxy mergers, so that it becomes
increasingly difficult to model the galaxy population at increasing
redshift by evolving the currently existing galaxies backward.

``Forward evolution,'' the approach we follow, is based on assuming a
fundamental theory of cosmology and galaxy formation.  A decade ago
this may have appeared to be a highly speculative approach, but it has
now become much more plausible since we now know that all available
data on large and intermediate scales is in spectacular agreement with
the LCDM model.  And as Fig. 1 shows, a modern LCDM-based SAM can
account very successfully for the nearby luminosity density.
Reasonable modifications (Somerville et al., in prep.) can also allow
SAM models to account for detailed features of the galaxy population,
such as the bimodality in galaxy colors and other properties observed
at both low redshift (e.g. Kauffmann et al. 2003) and out at least to
redshift $z\approx 1$ (e.g. Bell et al. 2004).  However, SAM models
and simulations have not yet been able to account for the bright
far-IR galaxies observed by SCUBA and other instruments including ISO
and Spitzer.\footnote{A recent paper (Baugh et al. 2004) presents a
SAM which can account for SCUBA sources by assuming that starbursts
have a ``top-heavy'' stellar initial mass function, in which most of
the stellar mass goes into high-mass stars.  This would of course
imply a very high supernova rate, and it remains to be seen whether
this is consistent with observed properties of such galaxies.}  This
is shown by Figures 7 and 8.  Figure 7 shows that three currently
popular SAM approaches and the latest hydrodynamical simulations
(Hernquist \& Springel 2003) both lead to predictions for the
evolution of the density of star formation vs. redshift which agree
with the optical data but not with the star formation indicated by the
SCUBA observations at 850 $\mu$m at $z\approx 2$ and the more
uncertain observations at $z\approx 4$.\footnote{See e.g. Chary \&
Elbaz (2001) for a similar conclusion regarding the star formation
indicated by the ISO 15 $\mu$m data at $z\approx 0.7$ (cf. Chary et
al. 2004).}  Figure 8 shows that the discrepancy between our SAM
predictions for number counts and observations is worst for the bright
SCUBA data.  Thus, while our SAM predictions appear to be in good
agreement with optical data both nearby and at high redshift, we are
clearly underestimating the contribution of galaxies at $z \gsim 0.7$
to the far-IR and perhaps also the mid-IR.  It follows that our
attenuation curves in Figs. 3 and 5 are likely to be reliable for
gamma-ray energies $\lsim 1$ TeV, but they probably underestimate the
attenuation at higher gamma-ray energies.

\section{Prospect}

We are now writing a series of papers on results from our current SAM
models, including revised versions of the EBL and gamma-ray
attenuation results presented here.  T. J. Cox and Patrik Jonsson, who
finished their PhDs with Primack in September 2004 (Cox 2004, Jonsson
2004), have been doing an extensive program of high-resolution
hydrodynamic simulations of galaxy interactions including the effects
of dust absorption, scattering, and reradiation.  On a longer
timescale, but still within approximately the next year, we plan to
incorporate these new insights into improved SAM models which we hope
will do a better job than our current ones of accounting for the
bright far-IR emission from galaxies.

\section{Dark Matter Annihilation at the Milky Way Center?}

In this last section, we turn to a different topic.  In a recent
paper, Gnedin \& Primack (2004) considered the effect of the
scattering of dark matter particles by the star cluster surrounding
the supermassive black hole at the center of the Galaxy.  They showed
using a Fokker-Planck treatment that this results in a unique radial
density profile proportional to $r^{-3/2}$, which will extend from
$\sim 10^{-4}$ pc from the black hole (where the density is cut off
due to annihilation of the dark matter, assumed to be weakly
interacting massive particles, WIMPs) to $\sim 2$ pc.  Since the
annihilation rate is proportional to the square of the dark matter
density, the cuspy density profile implies that the annihilation peaks
at the location of the black hole.  The high-energy gamma-ray signal
from the Galactic center observed by H.E.S.S. (Aharonian et al. 2004,
Horns et al. 2004) appears to have this sort of point-like character.
This interpretation of the signal implies a WIMP mass of $\sim 20$
TeV, which is much higher than was expected on the basis of common
supersymmetric models, but not obviously inconsistent.  WIMP
annihilation implies a characteristic cut off in the energy spectrum,
and the data taken in 2004 with the full four-telescope H.E.S.S. array
may also have adequate angular resolution to help discriminate this
interpretation from alternative ones.  Unfortunately, while the radial
profile of the central dark matter density is known, its magnitude is
not -- since there are processes that enhance the density and other
processes that diminish it, and none of these processes are understood
well enough to permit reliable calculations.  Thus if dark matter
annihilation is not discovered at the Galactic center the implications
for WIMPs may not be clear.


\begin{theacknowledgments}
Primack acknowledges support from NASA and NSF grants at UCSC.
Bullock has been supported as a Hubble Fellow at the CfA, and
Somerville as a staff member at STScI.
\end{theacknowledgments}



\newenvironment{reflist}{\begin{list}{}{\listparindent -0.20in
 \leftmargin 0.2in} \item \ \vspace{-.35in} }{\end{list}}

\vfill \eject

\begin{reflist}

\section{References}
\

Aharonian, F., et al. 2003, A\&A, 403, 523.

Aharonian, F., et al. 2004, A\&A, 425L, 13.

Baugh, C. M., et al. 2004, astro-ph/0406069.

Bell, E. F. 2004, ApJ, 608, 752.

Blanton, M. R., et al. 2003, ApJ, 592, 819  and ApJ, 594..186

Blain, A. W., Smail, I., Ivison, R. J., \& Knieb, J.-P. 1999, MNRAS,
302, 632.

Blumenthal, G. R., Faber, S. M.,  Primack, J. R., \& Rees, M. 1984,
Nature, 311, 517.

Chary, R., \& Elbaz, D. 2001, ApJ, 556, 562.

Chary, R., et al. 2004, ApJS, 154, 80.

Cole, S., Lacey, C. G.; Baugh, C. M.; Frenk, C. S. 2000, MNRAS, 319, 168.

Cole, S., et al. 2001, MNRAS, 326, 255

Cooray, A., et al. 2004, ApJ, 606, 611

Costamante, L., Aharonian, F., Ghisellini, G., \& Horns, D. 2004a, 
New Astronomy Reviews, 47, 677.

Costamante, L., Aharonian, F., Horns, D., \& Ghisellini, G. 2004b,
New Astronomy Reviews, 48, 469.

Cox, T. J. 2004, UCSC PhD dissertation,
\url{http://physics.ucsc.edu/~tj/work/thesis/}

De Lucia, G., Kauffmann, G., \& White, S. D. M. 2004, 349, 1101.

Elbaz, D., et al. 2002, A\&A, 384, 848.

Giavalisco, M., et al. 2004, ApJ, 600, L103.

Gnedin, O., \& Primack, J. R. 2004, Phys. Rev. Lett., 93, 061302.

Hernquist, L. \& Springel, V. 2003, MNRAS, 341, 1253.

Horns, D. 2004, Phys. Lett. B, submitted, astro-ph/0408192.

Jonsson, P. 2004, UCSC PhD dissertation,
\url{http://sunrise.familjenjonsson.org/thesis}

Kauffmann, G., et al. 2003, MNRAS, 341, 54.

Kauffmann, G., White, S. D. M., \& Guiderdoni, B. 1993, ApJ, 264, 201.

MacMinn, D., \& Primack, J. R. 1996, in {\it TeV Gamma Ray
Astrophysics}, ed. Heinz V\"olk and F. Aharonian, {\sl Space Science
Reviews}, 75, 413.

Magliocchetti, M., Salvaterra, R., \& Ferrara, A. 2003, MNRAS, 342,
L25

Matsumoto et al. 2000, in {\it ISO Surveys of a Dusty Universe},
ed. D. Lemke et al. (Springer Lecture Notes in Physics) 548, 96

Norberg, P., et al. 2002, MNRAS, 332, 827

Press, W. H., \& Schechter, P. 1974, ApJ, 187, 425.

Primack, J. R. 2004, in {\it Proceedings of 6th UCLA Symposium on Sources
and Detection of Dark Matter in the Universe}, Marina del Rey,
February 2004, ed. D. Cline, in press, astro-ph/0408359.

Primack, J. R., Bullock, J. S., Somerville, R. S., \& MacMinn, D. 1999,
{\it VERITAS Workshop on TeV Astrophysics of Extragalactic
Sources}, eds. M. Catanese \& T. Weekes, Astroparticle Physics, 11,
93.

Primack, J. R., Somerville, R. S., Bullock, J. S., \& Devriendt,
J. E. G. 2001, in {\it High Energy Gamma Ray Astronomy},
Proceedings of the International Symposium Gamma-2000, Heidelberg,
June 2000, eds. F. Aharonian and H. V\"olk, AIP Conf. Proc.,
558, 463.

Salvaterra, R., \& Ferrara, A. 2003, MNRAS, 340, L17

Seljak, U., et al. 2004, astro-ph/0407372.

Somerville, R. S. 2004, in Proceedings of the ESO/USM/MPE Workshop
{\it Multiwavelength Mapping of Galaxy Formation and Evolution},
eds. R. Bender \& A. Renzini, in press, astro-ph/0401570.

Somerville, R. S., Bullock, J. S., \& Livio M. 2003, ApJ, 593, 616.

Somerville, R. S., \& Kolatt, T. 1999, MNRAS, 305, 1.

Somerville, R. S., \& Primack, J. R. 1999, MNRAS, 310, 1087.

Somerville, R. S., Primack, J. R., \& Faber, S. M. 2001, MNRAS, 320, 504.

Spergel, D. N., et al. 2003, ApJS, 148, 175.

Tegmark, M., et al. 2004 PhysRevD.69.103501

Wechsler, R. S., Bullock, J. S., Primack, J. R., Kravtsov, A. V., \&
Dekel, A. 2002, ApJ, 581, 799.

White, S. D. M., \& Frenk, C. 1991, ApJ, 379, 52.

Wright, E. L. 2004, New Astron. Rev., 48, 465

\end{reflist}

\end{document}